\documentclass[aps, prd, twocolumn,a4paper]{revtex4}
\usepackage{ulem}
\usepackage{color}
\usepackage{float}
\usepackage{graphicx}
\usepackage{epsfig}
\usepackage{epstopdf}
\usepackage{amsmath}
\usepackage{subfig}
\usepackage[dvipsnames]{xcolor}
\newcommand{\be}{\begin{equation}}
\newcommand{\ee}{\end{equation}}
\newcommand{\ba}{\begin{eqnarray}}
\newcommand{\ea}{\end{eqnarray}}
\newcommand{\nn}{\nonumber\\}
\usepackage{graphicx}
\usepackage{color}
\begin{document}
\title{Covariant kinetic theory for effective fugacity quasi particle model and first order transport coefficients for hot QCD matter}
\author{Sukanya Mitra}
\email{sukanyam@iitgn.ac.in}
\author{Vinod Chandra}
\email{vchandra@iitgn.ac.in}
\affiliation{Indian Institute of Technology Gandhinagar, Palaj, Gandhinagar-382355, Gujarat, India}

\begin{abstract}
An effective relativistic  kinetic theory has been constructed for an interacting system of quarks, anti-quarks and gluons within a 
quasi-particle description of hot QCD medium at finite temperature and baryon chemical potential, where the interactions are encoded 
in the gluon and quark effective fugacities with non-trivial energy dispersions. The local conservations of stress-energy tensor and number 
current require the introduction of a mean field term in the transport equation which produces non-vanishing contribution to the first 
order transport coefficients. Such contribution has been observed to be significant for the temperatures which are closer to the QCD 
transition temperature, however, induces negligible contributions beyond a few times the transition temperature. As an implication, 
 impact of the mean field contribution on the the temperature dependence of  the shear viscosity, 
bulk viscosity  and thermal conductivity of a hot QCD medium in the presence of binary, elastic collisions among the constituents, has been investigated.
Visible effects have been observed for the temperature regime closer to the QCD transition temperature.
\\
\\

\end{abstract}
\maketitle

\section{Introduction}
 In view of the fact that heavy-ion experiments at relativistic heavy-ion collider (RHIC) and large hadron collider (LHC) 
 have already realized strongly coupled quark-gluon plasma (QGP) 
\cite{STAR,PHENIX,PHOBOS,ALICE}, interacting hot QCD equations of state (EOSs)  computed either  within the
 lattice QCD framework~ \cite{Lattice1,Lattice2,Lattice3,Lattice4} 
or the  improved Hard Thermal Loop(HTL)  perturbation theory up to three loops~\cite{HTLpt}, might play  important roles in modeling the equilibrium/isotropic state of the QGP.  
On the other hand, effective transport theory approaches beyond hot QCD transition temperature (weak coupling domain) have already shown their 
usefulness in understanding the bulk and the transport properties of the QGP/hot QCD matter \cite{AMY1,AMY2,Chen,Arnold,Xu}. These approaches 
not only require the microscopic definitions of various thermodynamic quantities for the QGP but also the appropriate momentum distributions as the 
inputs. To that end, mapping hot QCD equation of state(EOS) effects  in a system of  
effective gluons and quark-antiquarks (quasi-particles) with non-trivial dispersion relations~\cite{Peshier,Rebhan,Bluhm1,Thaler,Szabo,Bannur}, has turned out to be a 
viable approach in developing  covariant transport theory. Moreover, the effective kinetic equation is needed to obtain the first and second order dissipative hydrodynamic 
equations that depict the fluid-dynamic evolution of the QGP medium in addition to the determination of the first and the second order transport 
coefficients itself.

In this work,  we are presenting the foundations of a relativistic kinetic theory of many particle, multi-component systems, that effectively
represent the partonic interactions within the system through a quasi particle model, {\it viz.},  effective fugacity quasi particle model (EQPM)\cite{Chandra1,Chandra2,Chandra3}.
The EQPM has been constructed on the idea of mapping the hot QCD medium effects present in the EOSs of the strongly interacting system, 
created in the heavy ion collision experiments in terms of quasi-gluons and quasi-quarks/antiquarks with respective temperature dependent 
effective fugacity parameters. The temperature dependence of  the effective fugacities has been determinded
from the recent (2+1)-flavor lattice data of HotQCD Collaboration \cite{Lattice2}, realizing the medium as an effective Grand canonical system of 
these quasi-particles. Further, the EQPM at higher temperature (much beyond QCD transition temperature, $T_c$) approaches 
to the perturbative QCD as far as the effective coupling or the Debye mass are concerned.

The key finding of the present article is to identify the presence of mean field terms that is necessary for the conservation of particle number and energy
momentum tensor from a covariant kinetic equation in terms of its appropriate moments. In our analysis, 
we have observed that, the mean field term turns out to be dependent on the medium modified part of the energy dispersions for the 
effective gluons and quark-antiquarks . Treating the above mentioned
function as the force term in the relativistic transport equation and expressing the thermodynamic quantities in terms of the quasi particle
four-momenta, we conveniently obtain the conservation relations for particle current and energy-momentum under the EQPM. Following the conservation relations 
we can further achieve all the equilibrium thermodynamic laws for a first order hydrodynamic theory. Under this scheme, a complete formalism 
for estimation of the first order transport coefficients, that quantifies the thermal and viscous dissipations in a strongly interacting medium 
can be developed consistently, preserving the quasi particle excitations in the transport theory of the system.  

 It is to be noted, that the presence of the mean field terms in the effective kinetic theory with quasi-particle models based on 
the temperature dependent effective masses in the hot QCD medium has long been realized in the context of conservation laws from kinetic theory
\cite{Jeon,Dusling,Gorenstein} along with an explanation on the  fundamental reason for the presence  of such mean field terms. As mentioned 
earlier, this modifies the kinetic theory (microscopic) definition of the energy-momentum tensor so that the hot QCD thermodynamics could exactly 
be reproduced from the quasi-particle model realizing hot QCD as an effective Grand-canonical ensemble of effective gluon and  quark-antiquark 
degrees of freedom.These aspects are crucial while computing transport coefficients for the hot QCD/QGP medium along with deriving hydrodynamic 
equations from covariant kinetic theory including second and higher order relativistic dissipative hydrodynamic evolution equations
\cite{Jeon,Dusling,Tinti,Bluhm2,Kapusta1,Kapusta2,Greco1,Greco2}. In the context of effective mass quasi-particle model, dissipative hydrodynamics 
with and without anisotropy has already been constructed and the predictions are tested against the experimental observation \cite{Alqahtani1,Alqahtani2}. 
Here, in the context of the EQPM, an effective kinetic theory is constructed with appropriate form of the energy-momentum tensor with mean field contribution. 
The impact of mean field contributions to first order transport coefficients such as shear and bulk viscosities and thermal conductivity of 
a hot QCD medium with binary, elastic collisions among the effective gluons and quarks-antiquarks has been presented in the current manuscript. 
The derivation for second and third order dissipative hydrodynamics is beyond the scope of the present work.

The manuscript is organized as follows. Section II deals with the details of the EQPM model, the development of the effective kinetic theory 
and hydrodynamics under it and its application for estimating the viscous coefficients and thermal conductivity for a QGP system. Section III 
presents the results, depicting the significance of the mean field term on the temperature dependence of the transport coefficients. The 
article ends with a conclusion and outlook section, summarizing the relevance and details of the work and with a discussion about the possible
open horizons in this direction. 

\section{Formalism}
This  section consists of the theoretical set up required to construct a complete, many particle effective theory that
follows the EQPM consistently and hence the estimations of relevant transport parameters.

\subsection{Effective fugacity quasi particle model}
As mentioned earlier, the EQPM maps the hot QCD medium effects in to medium consist of non-interacting/weakly interacting quasi-gluons and quasi-quarks 
possessing the following form for their  equilibrium momentum distribution functions:

\begin{equation}
f^0_{g,q}=\frac{z_{g,q} \exp\big\{-\frac{ E_{p}}{T}\big\})}{1\mp z_{g,q} \exp\big\{-\frac{ E_{p}}{T}\big\})}.
\label{dist1}
\end{equation}
Here, $T$ is the temperature of the system and $E_p$ simply denotes the energy of a single bare parton, which for a gluon becomes $E_p=|\vec{p}|$ 
and for a quark turns out to be $E_p=\sqrt{|\vec{p}|^2+m_q^2}$, with $m_q$ as the quark mass. This model can be straightforwardly extended to
include finite baryon chemical potential in the quark/anti-quark equilibrium distribution function in the following way,

\begin{equation}
f^0_{q,\bar{q}}=\frac{z_{g,q} \exp\big\{-\frac{ E_{p}\mp \mu_q}{T}\big\})}{1\mp z_{g,q} \exp\big\{-\frac{ E_{p}\mp\mu_q}{T}\big\})}.
\label{dist2}
\end{equation}

The local equilibrium can be straightforwardly described simply by generalizing Eq.(\ref{dist1}) in the co-moving frame of the fluid, 
defined by the hydrodynamic four-velocity $u^{\mu}=(1,0)$ in the local rest frame (LRF)  as, 
\begin{equation}
f_{g,q}=\frac{z_{g,q} \exp\big\{-\frac{ u^\mu p_\mu}{T}\big\}}{1\mp z_{g,q} \exp\big\{-\frac{  u^\mu p_\mu}{T}\big\}}.
\label{dist3}
\end{equation}

Here, we define $p^{\mu}=(E_{p},\vec{p})$ is the bare four-momenta (without including the effects of interactions)
and $\tilde{p}^{\mu}=(\omega_p,\vec{p})$ is the quasi-particle four-momenta under the EQPM, corresponding to a parton. The three momenta 
$\vec{p}$ is not altered under EQPM, where the single particle energy has been modified via a dispersion relation as follows,

\begin{equation}
 \omega_p=E_{p}+\delta \omega,~~~~~~~~~~~~~~~~\delta\omega=T^2 \partial_{T}ln{z_{g,q}}~,
 \label{Eq1}
 \end{equation}
 with $z_{g,q}$ is the fugacity parameter for gluons and quarks respectively, through which the interactions are being mapped into Eq.(\ref{dist1}).
$\delta\omega(T)$ is a pure temperature $(T(x))$ dependent quantity which is again function of four space-time coordinate $x^{\mu}\equiv (t,\vec{x})$.
For a massless case with gluons and light quarks, Eq.(\ref{Eq1}) simply reduces to,
\begin{equation}
 \omega_p= |{\vec{p}}| +\delta \omega~.
 \label{Eq2}
\end{equation}
In the light of above discussion the quasi particle and bare particle four momenta can be related in a local rest frame
as follows,
\begin{equation}
 \tilde{p}^{\mu}={p}^{\mu}+\delta\omega\ u^{\mu}~,
 \label{Eq3}
\end{equation}
which picks up modification of only the energy (zeroth) component  of particle 4-momenta through the dispersion relation (\ref{Eq1}).

\subsection{Fundamental quantities of effective kinetic theory under EQPM}

In order to set up a covariant kinetic theory for a many particle, multi-component system, under the assumptions of EQPM mentioned above,
we first need to define the basic macroscopic quantities that describe the thermodynamic state of the system. We start with the particle
4-flow which manifests the particle number density $n(x)$ and particle current $\vec{j}(x)$ as its zeroth and $i^{th}$ component.
The quasi particle four flow $N^{\mu}(x)$ can be defined in terms of bare momenta as the following,
\begin{equation}
 N^{\mu}(x)=\sum_{k=1}^{N}\nu_k\int\frac{d^{3}|\vec{p_k}|}{(2\pi)^3}\frac{p_{k}^{\mu}}{E_{p_k}}f_{k}(x,p_k)~,
 \label{Eq4}
\end{equation}
that retains the expression of particle number density $n(x)$ under EQPM as the following,
\begin{equation}
 n(x)=N^{\mu}u_{\mu}=\sum_{k=1}^{N}\nu_k\int\frac{d^{3}|\vec{p_k}|}{(2\pi)^3} f_k(x,p_k)~.
 \label{Eq5}
\end{equation}
Here $f_k(x,p_k)$ is the single particle momentum distribution belonging to $k^{th}$ species, that is a function of space-time coordinate 
and particle momenta and $\nu_k$ is the corresponding degeneracy factor. Throughout the analysis, the subscript $k$ denotes the particle species. 
Now, it can be shown that $N^{\mu}$ can be  expressed in terms of dressed momenta $\tilde{p}$ as follows, 

\begin{eqnarray}
 N^{\mu}(x)=&&\sum_{k=1}^{N}\nu_k\int\frac{d^{3}|\vec{\tilde{p_k}}|}{(2\pi)^3{{\omega_{p}}_k}}{{\tilde{p}}_k^{\mu}}f_k(x,{\tilde{p}}_k)\nonumber\\
           +&&\delta\omega \sum_{k=1}^{N}\nu_k\int\frac{d^{3}|\vec{\tilde{p_k}}|}{(2\pi)^3{\omega_{{p}_{k}}}}\frac{\langle{{\tilde{p}}_{k}^{\mu}}\rangle}
           {|\vec{\tilde{p_k}}|}f_k(x,\tilde{p_k})~.
\label{Eq6}
 \end{eqnarray}

Here $\langle{\tilde{p}^{\mu}}\rangle=\Delta^{\mu\nu}\tilde{p}_{\nu}$ is the irreducible tensor of rank one, with $\Delta^{\mu\nu}=g^{\mu\nu}-u^{\mu}u^{\nu}$
as the projection operator. Throughout the analysis the metric $g^{\mu\nu}$ has taken to be $g^{\mu\nu}=(1,-1,-1,-1)$.
The identical individual components of $N^{\mu}$ from Eq.(\ref{Eq4}) and (\ref{Eq6}) and the unaltered form of $n$ as obtained from Eq.(\ref{Eq5}),
confirms the expression of $N^{\mu}$ as given by Eq.(\ref{Eq6}) in terms of dressed momenta $\tilde{p}^{\mu}$.

Next, we focus on the energy momentum tensor $T^{\mu\nu}(x)$ whose different components describes the energy density and momentum flow.
The quasi particle energy-momentum tensor $T^{\mu\nu}(x)$ can be defined under EQPM in terms of bare momenta as the following,

\begin{eqnarray}
 T^{\mu\nu}(x)=&&\sum_{k=1}^{N}\nu_k\int\frac{d^{3}|\vec{p}_{k}|}{(2\pi)^3E_{p_{k}}}p_k^{\mu}p_k^{\nu}f_k(x,p_k)\nonumber\\
              +&&\delta\omega u^{\mu} u^{\nu}\sum_{k=1}^{N}\nu_k\int\frac{d^{3}|\vec{p_k}|}{(2\pi)^3}f_k(x,p_k)~.
\label{Eq7}
\end{eqnarray}

Note that Eq. (\ref{Eq7})  gives the expression of quasi particle energy density and pressure respectively as,
\begin{eqnarray}
 \epsilon(x)=&& u_{\mu}u_{\nu}T^{\mu\nu}\nonumber\\
            =&&\sum_{k=1}^{N}\nu_k\int\frac{d^{3}|\vec{p_k}|}{(2\pi)^3} \omega_{p_k} f_k(x,p_k)~,
 \label{Eq8}
 \\
 P(x)=&&-\frac{1}{3}\Delta_{\mu\nu}T^{\mu\nu}\nonumber\\
     =&&\frac{1}{3}\nu_k\int\frac{d^{3}|\vec{p_k}|}{(2\pi)^3} |\vec{p_k}| f_k(x,p_k)~.
 \label{Eq9}
\end{eqnarray}

In terms of dressed momenta, $T^{\mu\nu}$ can be shown to take the form,

\begin{eqnarray}
 T^{\mu\nu}(x)=&&\sum_{k=1}^{N}\nu_k\int\frac{d^{3}|\vec{\tilde{p_k}}|}{(2\pi)^3{\omega_{p_k}}}{\tilde{p_k}^{\mu}}{\tilde{p_k}^{\nu}}f_k(x,\tilde{p_k})\nonumber\\
 +&& \delta\omega \sum_{k=1}^{N} \nu_k\int\frac{d^{3}|\vec{\tilde{p_k}}|}{(2\pi)^3{\omega_{p_k}}}\frac{\langle{\tilde{p_k}^{\mu}}{\tilde{p_k}^{\nu}}\rangle}
 {|\vec{\tilde{p_k}}|}f_k(x,\tilde{p_k})~,\nonumber\\
 \label{Eq10}
\end{eqnarray}
with 
${\langle{\tilde{p}^{\mu}}{\tilde{p}^{\nu}}\rangle}=\frac{1}{2}\big\{ \Delta^{\mu\alpha}\Delta^{\nu\beta}+\Delta^{\mu\beta}\Delta^{\nu\alpha} \big\}\tilde{p}_{\alpha}\tilde{p}_{\beta}$
as the irreducible tensor of rank two.
Eq.(\ref{Eq10}) readily traces back the expression of $\epsilon$ and $P$ as given by Eq.(\ref{Eq8}) and (\ref{Eq9}).

Finally, we provide the microscopic definition of entropy 4-current as,
\begin{equation}
 S^{\mu}=-\sum_{k=1}^{N}\nu_k\int\frac{d^3|\vec{\tilde{p_k}}|}{(2\pi)^3\omega_{p_k}}\tilde{p}_{k}^{\mu}\big\{f_k lnf_k\mp(1\pm f_k)ln(1\pm f_k)\big\}~. 
 \label{Eq55}
 \end{equation}
Contraction of Eq.(\ref{Eq55}) with $u^{\mu}$ gives the entropy density as follows,
\begin{equation}
 s=S^{\mu}u_{\mu}=-\sum_{k=1}^{N}\nu_k\int\frac{d^3|\vec{\tilde{p_k}}|}{(2\pi)^3} \big\{f_k lnf_k\mp(1\pm f_k)ln(1\pm f_k)\big\}~.
 \label{Eq56}
\end{equation}

\subsection{Conservation laws}

We start with the relativistic transport equation of the single quasi-particle distribution function, that can be given by the following covariant equation,
\begin{eqnarray}
 \frac{1}{\omega_{p_k}} \tilde p_{k}^{\mu}\partial_{\mu}f_k(x,\tilde{p_k})+\vec{F}\cdot\vec{\nabla}_{p_{k}} f_{k}= \sum_{l=1}^{N}C_{kl}[f_{k},f_{l}]~,\nonumber\\
 ~~~~~~~~~~[k=1,....,N]
 \label{Eq12}
\end{eqnarray}
with $\vec{F}$ as the external force and $C_{kl}$ as the collision integral given by,

\begin{eqnarray}
 &&C_{kl}[f_{k},f_{l}]=\frac{1}{2} \frac{\nu_{l}}{2\omega_{p_k}} \int d\Gamma_{\tilde{p}^{}_{l}} d\Gamma_{\tilde{p}'_{k}}  d\Gamma_{\tilde{p}'_{l}}  
          \delta^{4}(\tilde{p}_{k}+\tilde{p}_{l}-\tilde{p}'_{k}-\tilde{p}'_{l})  \nn &&\times  (2\pi)^4[f_{k}(\tilde{p}'_{k}) f_{l}(\tilde{p}'_{l}) 
          \{1\pm f_{k}(\tilde{p}_{k})\}\{1\pm f_{l}(\tilde{p}_{l})\}\nn&&-
          f_{k}(\tilde{p}_{k})f_{l}(\tilde{p}_{l})\{1\pm f_{k}(\tilde{p}'_{k})\}\{1\pm f_{l}(\tilde{p}'_{l})\}]\nn && \times \langle|M_{k+l\rightarrow k+l}|^{2}\rangle~.
\label{eq-R2}     
\end{eqnarray}
The phase space factor is given by the notation $d\Gamma_{\tilde{p}_{i}}=\frac{d^3 \vec {\tilde{p}}_{i} }{(2\pi)^3 2\omega_i}$.
The overall,  $\frac{1}{2}$ factor appears due to the symmetry in order to compensate for the double counting of
final states that occurs by interchanging $\tilde{p}'_{k}$ and $\tilde{p}'_{l}$. $\nu_{l}$ is the degeneracy of $2^{nd}$ particle 
that belongs to $l^{th}$ species. $\langle|M_{k+l\rightarrow k+l}|^{2}\rangle$ is the QCD scattering amplitudes for $2\rightarrow2$ binary, 
elastic processes are taken from ~\cite{Combridge}, which are averaged over the spin and color degrees of freedom of the initial states 
and summed over the final states. However, the inelastic processes like $q\overline{q}\rightarrow gg$, have been ignored in the present
case, because of the fact that  they do not have a forward peak in the differential cross section and thus their contributions will presumably be 
much smaller compared to the elastic ones.

\subsubsection{Conservation of particle current}

Integrating both sides of Eq.(\ref{Eq12}) over $\int\frac{d^3|\vec{\tilde{p_k}}|}{(2\pi)^3}$ and summing over $k=[0,N]$ we obtain,
\begin{equation}
 \sum_{k=1}^{N}\int\frac{d^3|\vec{\tilde{p_k}}|}{(2\pi)^3} \frac{1}{\omega_{p_k}} \tilde p_{k}^{\mu}\partial_{\mu}f_{k}(x,\tilde{p_{k}})+
 \sum_{k=1}^{N}\int\frac{d^3|\vec{\tilde{p_{k}}}|}{(2\pi)^3} F^{i}\frac{\partial f_k}{\partial p_{k}^{i}}=0~.
 \label{Eq13}
\end{equation}

The right hand side of Eq.(\ref{Eq13}) is zero by the virtue of zeroth moment of summation invariance.
Now we define the force term as,
\begin{equation}
 F^{i}=-\partial_{\mu}\big\{\delta\omega \ u^{\mu} u^{i} \big\}~.
 \label{Eq14}
\end{equation}
With this form of $F^{i}$, the integration on the second term of left hand side of Eq.(\ref{Eq13}) also reduces to zero.
Now following the definition of particle 4-flow from Eq.(\ref{Eq6}) and performing the necessary integrations of Eq.(\ref{Eq13}),
we can achieve the conservation of the particle flow,

\begin{equation}
 \partial_{\mu} N^{\mu}=0~,
 \label{Eq15}
\end{equation}

where the momentum integration over the second term of $N^{\mu}$ from Eq.(\ref{Eq6}) and over 
$\int\frac{d^3|\vec{\tilde{p_k}}|}{(2\pi)^3}\partial\big\{{\frac{\tilde{p_k}^{\mu}}{\omega_{p_k}}}\big\}f_{k}$ exactly cancels each other to preserve
particle flow conservation.

\subsubsection{Conservation of energy momentum}
Integrating both sides of Eq.(\ref{Eq12}) over $\int\frac{d^3|\vec{\tilde{p_k}}|}{(2\pi)^3}\tilde{p_k}^{\nu}$ and summing over $k$ we now obtain,

\begin{eqnarray}
 &&\sum_{k=1}^{N}\int\frac{d^3|\vec{\tilde{p_{k}}}|}{(2\pi)^3} \frac{1}{\omega_{p_k}} \tilde p_k^{\mu}\tilde p_k^{\nu}\partial_{\mu}f_k(x,\tilde{p_k})\nonumber\\
 &&+\sum_{k=1}^{N}\int\frac{d^3|\vec{\tilde{p_k}}|}{(2\pi)^3} \tilde p_{k}^{\nu} F^{i}\frac{\partial f_k}{\partial p_{k}^{i}}=0
 \label{Eq16}
\end{eqnarray}

This time the right hand side of Eq.(\ref{Eq16}) is zero by the virtue of first moment of summation invariance.

Defining the force term as Eq.(\ref{Eq14}) and adopting the definition of quasi particle stress energy tensor from Eq.(\ref{Eq10}),
the space-time derivative over $T^{\mu\nu}$ can be written as,

\begin{eqnarray}
 \partial_{\mu}T^{\mu\nu}=&&
 \partial_{\mu}\bigg\{\delta\omega \sum_{k=1}^{N} \nu_k\int\frac{d^{3}|\vec{\tilde{p_k}}|}{(2\pi)^3 \omega_{p_k}}\frac{\langle\tilde{p_k}^{\mu}\tilde{p_k}^{\nu}\rangle}{|\vec{\tilde{p_k}}|}
 f_k(x,\tilde{p_k})\bigg\}\nonumber\\
 +&&\sum_{k=1}^{N}\nu_k\int\frac{d^{3}|\vec{\tilde{p_k}}|}{(2\pi)^3 }\partial_{\mu}\big\{\frac{\tilde{p_k}^{\mu}\tilde{p_k}^{\nu}}{\omega_{p_k}}\big\}f_k(x,\tilde{p_k})\nonumber\\
 -&&\nu_kn\partial_{\mu}\big\{\delta\omega \  u^{\mu} u^{\nu}\big\}~,
 \label{Eq17}
\end{eqnarray}

where the integration over the force term simply reduces to a momentum independent quantity $n\partial_{\mu}\big\{\delta\omega \ u^{\mu} u^{\nu}\big\}$.
The space-time derivative over $f_k$ multiplied with $\delta\omega$ has been ignored considering $\frac{\delta\omega}{\langle|\vec{\tilde{p}}|\rangle_{T}}<<1$,
where $\langle ~ \rangle_{T}$ stands for the notation of thermal average.
It can be shown that the addition of first and the second term on the right hand side of Eq.(\ref{Eq17}) exactly cancels the force term to make the 
right hand side zero whole together, leading to

\begin{equation}
 \partial_{\mu}T^{\mu\nu}=0~.
 \label{Eq18}
\end{equation}

We must notice that the partial derivative on the first term of left hand side of Eq.(\ref{Eq16}) can not be simply taken outside of the 
integral, since now the phase space factor $\int\frac{d^{3}|\vec{\tilde{p_k}}|}{(2\pi)^3 \omega_{p_k}}$ contains a space-time dependent
quasi-particle energy $\omega_{p_k}$ given by Eq.(\ref{Eq2}). Hence the extraction of the partial derivative outside integral produces
an extra term $\int\frac{d^{3}|\vec{\tilde{p_k}}|}{(2\pi)^3 }\partial_{\mu}\big\{\frac{\tilde{p_{k}}^{\mu}\tilde{p_{k}}^{\nu}}{\omega_{p_k}}\big\}f_k(x,\tilde{p_k})$.
So from this current analysis we can conclude that the difference of this term and the force term generated due to the quasi particle
excitations as a function of dispersion parameter $\delta\omega$, produces the exact additional piece of quasi particle energy-momentum tensor over
the usual one ($\int\frac{d^{3}|\vec{p_k}|}{(2\pi)^3\omega_{p_k}}\tilde{p}_{k}^{\mu}\tilde{p}_{k}^{\nu}f(x,\tilde{p}_k)$) from conventional kinetic theory.
Eq.(\ref{Eq18}) gives the desired energy-momentum conservation under the EQPM scheme. 
Hence Eq.(\ref{Eq12}) with the force term given in Eq.(\ref{Eq14}) gives effective kinetic theory description of the interacting
partons in a thermal medium under EQPM scheme.

\subsection{Estimation of transport coefficients} 
\subsubsection{Solution to  relativistic transport equation}
Determination of the transport coefficients requires the knowledge of the system 
away from equilibrium. This could be done by first set-up the relativistic transport equation and look for the appropriate solutions. Here, 
in order to estimate the transport coefficients, one needs  solve the relativistic transport equation (\ref{Eq12}). For this purpose
we employ the Chapman-Enskog (CE) method, which is an iterative technique, where from the known lower order distribution function the 
unknown next order can be determined by successive approximation. Furthermore, to solve the transport equation Eq.(\ref{Eq12}) for $k^{th}$ 
species, we need to linearize the collision term on right hand side by introducing the relaxation time $\tau_{k}$ over the deviation part of
the next to leading order momentum distribution from the lowest order in following manner,

\begin{equation}
 \frac{1}{\omega_{p_{k}}} \tilde p_{k}^{\mu}\partial_{\mu}f_{k}^{0}(x,\tilde{p_{k}})+{F^{i}}\frac{\partial f^{0}_k}{\partial p^{i}_{k}}=-\frac{\delta f_{k}}{\tau_{k}}
 =-\frac{f_{k}^{0}(1\pm f_{k}^{0})\phi_{k}}{\tau_{k}}~.
 \label{Eq25}
\end{equation}
Clearly, $f_{k}^{0}$ provides the leading order momentum distribution which is the equilibrium distribution function and $\delta f_k$ accounts
for the correction to the next to leading order corresponding to $k^{th}$ species. Hence, $\phi_k$ denotes deviation of momentum distribution from its equilibrium
value that quantifies the dissipation in the medium.

Now we retrieve the definition of equilibrium distribution function of quasi-partons under the EQPM from Eq.(\ref{dist1}), (\ref{dist2}) and (\ref{dist3}).
In covariant notation with 4-momenta $\tilde{p_{k}}^{\mu}$ the above equations can be written as,

\begin{equation}
 f^{0}_{k}(x,\tilde{p}_{k})=\frac{1}{exp\big\{\frac{\tilde{p}^{\mu}_{k}u_{\mu}}{T}-\frac{\mu_{k}}{T}\big\}}~,
 \label{Eq24}
\end{equation}
with $\mu_k=\mu_{B_k}+\delta \omega +Tln z_{k}$, such that $\exp\{-\frac{\mu_{k}}{T}\}$ is the total effective
fugacity due to the baryon chemical potential and the quasi particle excitation effects.

Here we need to make an important remark. After defining the local equilibrium distribution function $f_{k}^{0}$, we
can identify $T$ and $\exp\{-\frac{\mu_{k}}{T}\}$, respectively as the temperature and effective fugacity of the system,
only after specifying the following conditions,

\begin{eqnarray}
 \sum_{k=1}^{N}\int\frac{d^{3}|\vec{\tilde{p_k}}|}{(2\pi)^3{\omega_{p_{k}}}} \tilde{p}^{\mu}_{k}u_{\mu}\delta f_{k}=0~,\\
 \sum_{k=1}^{N}\int\frac{d^{3}|\vec{\tilde{p_k}}|}{(2\pi)^3{\omega_{p_{k}}}} (\tilde{p}^{\mu}_{k}u_{\mu})^2\delta f_{k}=0~,
\end{eqnarray}
which follows from the fact that contraction of the non-equilibrium part of $N^{\mu}$ and $T^{\mu\nu}$ from equation
(\ref{Eq6}) and (\ref{Eq10}) with hydrodynamic velocity $u^{\mu}$, gives rise to zero (which is nothing but Landau-Lifshitz condition).
In such situation the expression of energy density and pressure from Eq.(\ref{Eq8}) and (\ref{Eq9}) contain only the 
equilibrium part of the distribution function. So we can conclude within the scheme of first order CE method, the 
particle number density and energy density can be specified by the equilibrium distribution function alone. Under such
circumstances, the equilibrium part of particle 4-flow and energy-momentum tensor can be expressed by the following macroscopic
definition,
\begin{eqnarray}
N^{\mu}=&&nu^{\mu}~.
\label{Eq6A}
\\
T^{\mu\nu}=&&\epsilon u^{\mu}u^{\nu}-P\Delta^{\mu\nu}~.
\label{Eq11}
\end{eqnarray}
Following same line of argument, the macroscopic definition of equilibrium entropy density becomes,
\begin{equation}
 s=\sum_{k=1}^{N}\bigg\{\frac{\epsilon_k + P_k}{T}-\frac{n_k\mu_k}{T}\bigg\}~.
 \label{Eq57}
\end{equation}

\subsubsection{Equilibrium thermodynamic laws}
From the thermodynamic definition of $N^{\mu}(x)$ and $T^{\mu\nu}(x)$ from Eq.(\ref{Eq6A}) and Eq.(\ref{Eq11}) respectively,
and following their conservation laws from Eq.(\ref{Eq15}) and Eq.(\ref{Eq18}) respectively, we can achieve the 
equilibrium thermodynamic laws of macroscopic state variable such as number density ($n$), energy per particle ($e$) and
hydrodynamic velocity $u^{\mu}(x)$, as follows,

\begin{eqnarray}
 Dn_k=&&-n_k \partial\cdot u~,
 \label{Eq19}
 \\
 \sum_{k=1}^{N}x_k De_k=&&-\bigg\{\frac{\sum_{k=1}^{N}P_k}{\sum_{k=1}^{N} n_k}\bigg\}\partial\cdot u~,
 \label{Eq20}
 \\
 Du^{\mu}=&&\frac{\nabla^{\mu}P}{nh}~,
 \label{21}
 \end{eqnarray}

with $D=u^{\mu}\partial_{\mu}\equiv\frac{\partial}{\partial t}$ as the convective time derivative and $h$ as enthalpy
per particle $h=e+\frac{P}{n}$ of the system. $P_k$ is the partial pressure belongs to
$k^{th}$ species that is related to total pressure as $P=\sum_{k=1}^{N}P_{k}$. $x_k=n_k/n$ denotes the particle fraction given by 
the ratio of particle number of $n^{th}$ species to total particle number, $x_k=n_k/n$. Following the prescription, the total
energy density can be given as, $\epsilon=\sum_{k=1}^{N}\epsilon_k=\sum_{k=1}^{N} e_k n_k$.

\subsubsection{Linearized solution of the deviation function}
Following the definition of equilibrium distribution function from Eq.(\ref{Eq24}), the second term on the left hand side of 
Eq.(\ref{Eq25}) vanishes for a co-moving frame, whereas the first term produces a number of terms containing thermodynamic 
forces giving rise to a number of transport processes as follows,

\begin{equation}
 Q_{k}X+\langle\tilde{p}_{k}^{\mu}\rangle(\omega_{p_{k}}-h_k)X_{q\mu}-\langle\tilde{p}_k^{\mu}\tilde{p}_k^{\nu}\rangle X_{\mu\nu}=-\frac{T\omega_{p_k}}{\tau_k}\phi_{k}~.
 \label{Eq26}
\end{equation}
with $Q_k=\frac{1}{3}\big\{|\vec{\tilde{p_k}}|^2-3\omega^2_{p_k}c_s^2\big\}$ where $c_s$ is the velocity of sound.
The thermodynamic forces such as the bulk viscous force, thermal force and shear viscous force are defined respectively
as follows,

\begin{eqnarray}
 X=&&\partial\cdot u~,
 \label{Eq27}\\
 X_{q}^{\mu}=&&\big\{\frac{\nabla^{\mu}T}{T}-\frac{\nabla^{\mu}P}{nh}\big\}~,
 \label{Eq28}\\
 X_{\mu\nu}=&&\langle\partial_{\mu}u_{\nu}\rangle~.
 \label{Eq29}
 \end{eqnarray}
Since thermodynamic forces are independent, in order to be a solution of Eq.(\ref{Eq26}), the deviation function
$\phi_{k}$ must be a linear combination of thermodynamic forces with a number of unknown coefficients,
\begin{equation}
 \phi_k=A_k X +B_{k}^{\mu}X_{q\mu}-C_{k}^{\mu\nu}X_{\mu\nu}~.
 \label{Eq30}
\end{equation}
The coefficients with proper tensorial ranks can be determined from Eq.(\ref{Eq26}) itself as the following,

\begin{eqnarray}
 A_{k}=&&\frac{Q_k}{\big\{-\frac{T\omega_{p_k}}{\tau_k}\big\}}~,
 \label{Eq31}\\
 B_{k}=&&\langle\tilde{p}_{k}^{\mu}\rangle\frac{(\omega_{p_k}-h_k)}{\big\{-\frac{T\omega_{p_k}}{\tau_k}\big\}}~,
 \label{Eq32}\\
 C_{k}^{\mu\nu}=&&\frac{\langle\tilde{p}_{k}^{\mu}\tilde{p}_{k}^{\nu}\rangle}{\big\{-\frac{T\omega_{p_k}}{\tau_k}\big\}}~.
 \label{Eq33}
\end{eqnarray}
Therefore, it cab be observed that through these coefficients which contain the thermal relaxation times of quasi-partons, the 
dynamic interactions of the medium enter in the expression of the deviation function, which are finally inserted in
the expressions of transport coefficients.

\subsubsection{Decomposition  of the energy-momentum tensor}
In order to decompose the energy momentum tensor in an equilibrium and an out of equilibrium part, we first define the pressure
tensor in the following way,

\begin{equation}
 P^{\mu\nu}=\Delta^{\mu}_{\sigma}T^{\sigma\tau}\Delta^{\nu}_{\tau}~.
 \label{Eq34}
\end{equation}

Following the covariant definition of $T^{\mu\nu}$ under the EQPM from Eq.(\ref{Eq10}), the pressure tensor
yields the form given below,

\begin{eqnarray}
 P^{\mu\nu}=&&\sum_{k=1}^{N}\nu_k\int\frac{d^{3}|\vec{\tilde{p_k}}|}{(2\pi)^3{\omega_{p_k}}}\langle{\tilde{p_k}^{\mu}}{\tilde{p_k}^{\nu}}\rangle f_k(x,\tilde{p_k})\nonumber\\
           +&& \delta\omega\ \sum_{k=1}^{N} \nu_k\int\frac{d^{3}|\vec{\tilde{p_k}}|}{(2\pi)^3{\omega_{p_k}}}\frac{\langle{\tilde{p_k}^{\mu}}{\tilde{p_k}^{\nu}}\rangle}
           {|\vec{\tilde{p_k}}|}f_k(x,\tilde{p_k})~.
 \label{Eq35}
\end{eqnarray}

In the LRF,  $P^{\mu\nu}$ is purely spatial,

\begin{eqnarray}
 P^{00}_{LRF}=&&P^{0i}_{LRF}=P^{i0}_{LRF}=0~,
 \label{Eq36}
 \\
 P^{ij}_{LRF}=&&\sum_{k=1}^{N}\nu_k\int\frac{d^{3}|\vec{\tilde{p_k}}|}{(2\pi)^3{\omega_{p_k}}}{\tilde{p_k}^{i}}{\tilde{p_k}^{j}} f_k(x,\tilde{p_k})\nonumber\\
             +&& \delta\omega\ \sum_{k=1}^{N}\nu_k\int\frac{d^{3}|\vec{\tilde{p_k}}|}{(2\pi)^3{\omega_{p_k}}}\frac{{\tilde{p_k}^{i}}{\tilde{p_k}^{j}}}{|\vec{\tilde{p_k}}|}f_k(x,\tilde{p_k})~.
 \label{Eq37}
\end{eqnarray}

Now we decompose $P^{\mu\nu}$ in a reversible and an irreversible part, that picks up respectively the equilibrium and non-equilibrium
components of $f_k$ in Eq.(\ref{Eq35}),

\begin{equation}
 P^{\mu\nu}=-P\Delta^{\mu\nu}+\Pi^{\mu\nu}~.
 \label{Eq38}
\end{equation}

The reversible part is addressed by the equilibrium distribution function $f_k^0$ as the following,
\begin{eqnarray}
 -P\Delta^{\mu\nu}=&&\sum_{k=1}^{N}\nu_k\int\frac{d^{3}|\vec{\tilde{p_k}}|}{(2\pi)^3{\omega_{p_k}}}\langle{\tilde{p_k}^{\mu}}{\tilde{p_k}^{\nu}}\rangle f_k^0(x,\tilde{p_k})\nonumber\\
                  +&& \delta\omega \ \sum_{k=1}^{N}\nu_k\int\frac{d^{3}|\vec{\tilde{p_k}}|}{(2\pi)^3{\omega_{p_k}}}\frac{\langle{\tilde{p_k}^{\mu}}{\tilde{p_k}^{\nu}}\rangle}
                  {|\vec{\tilde{p_k}}|}f_k^0(x,\tilde{p_k})~,\nonumber\\
 \label{Eq39}
\end{eqnarray}
which on contracting with $\Delta_{\mu\nu}$ simply leads to Eq.(\ref{Eq9}), revealing $P$ as the local hydrostatic pressure.
 
However, the irreversible part $\Pi^{\mu\nu}$, named by viscous pressure tensor, includes the non-equilibrium part of $f_k$ only,
leading to the following expression,

\begin{eqnarray}
 \Pi^{\mu\nu}=&&\sum_{k=1}^{N}\nu_k\int\frac{d^{3}|\vec{\tilde{p_k}}|}{(2\pi)^3{\omega_{p_k}}}\langle{\tilde{p_k}^{\mu}}{\tilde{p_k}^{\nu}}\rangle \delta f_k\nonumber\\
             +&& \delta\omega\ \sum_{k=1}^{N}\nu_k\int\frac{d^{3}|\vec{\tilde{p_k}}|}{(2\pi)^3{\omega_{p_k}}}\frac{\langle{\tilde{p_k}^{\mu}}{\tilde{p_k}^{\nu}}\rangle}{|\vec{\tilde{p_k}}|} \delta f_k~.
 \label{Eq40}
\end{eqnarray}

From Eq.(\ref{Eq40}), we can see that the viscous pressure tensor is orthogonal to hydrodynamic velocity,

\begin{equation}
 \Pi^{\mu\nu}u_{\mu}=0~.
 \label{Eq41}
\end{equation}

The heat flow is defined as the difference of energy flow and the flow of enthalpy carried by the particle,

\begin{equation}
 I_{q}^{\mu}=u_{\nu}T^{\nu\sigma}\Delta^{\mu}_{\sigma}-h N^{\sigma}\Delta^{\mu}_{\sigma}~.
 \label{Eq42}
\end{equation}

Putting expressions of $T^{\nu\sigma}$ and $N^{\sigma}$ from Eq.(\ref{Eq10}) and (\ref{Eq6}) respectively, it can be shown
that in the LRF the heat flow is purely spatial as well,

\begin{eqnarray}
 I_q^0=&&0~,
 \label{Eq43}
 \\
 I_q^i=&&T^{i0}_{LRF}-N_{LRF}^{i}~.
 \label{Eq44}
\end{eqnarray}

From Eq.(\ref{Eq42}), it is evident that $I_q^{\mu}$ is also orthogonal to $u^{\mu}$,

\begin{equation}
 I_q^{\mu}u_{\mu}=0~.
 \label{Eq45}
\end{equation}

From Eq.(\ref{Eq42}), it is also observed that heat flow only retains the non-equilibrium part of $f_k$, while the 
equilibrium $f_k^{0}$ produces zero contraction in heat flow, leading to the following expression,

\begin{eqnarray}
 I_q^{\mu}=&&u_{\nu}\Delta^{\mu}_{\sigma}\sum_{k=1}^{N}\nu_k\int\frac{d^{3}|\vec{\tilde{p_k}}|}{(2\pi)^3{\omega_{p_k}}}\tilde{p_k}^{\nu}\tilde{p_k}^{\sigma}\delta f_k\nonumber\\
-&&h \Delta^{\mu}_{\sigma} \bigg\{ \sum_{k=1}^{N}\nu_k\int\frac{d^{3}|\vec{\tilde{p_k}}|}{(2\pi)^3{\omega_{p_k}}}\tilde{p_k}^{\sigma} \delta f_k\nonumber\\ 
+&&\delta\omega \sum_{k=1}^{N} \nu_k\int\frac{d^{3}|\vec{\tilde{p_k}}|}{(2\pi)^3{\omega_{p_k}}} \frac{\langle \tilde{p_k}^{\sigma} \rangle}{|\vec{\tilde{p_k}}|} \delta f_k \bigg\}~.
\label{Eq46}
\end{eqnarray}

It is worth noting that in viscous pressure the additional term due to dispersion parameter $\delta\omega$ is contributed from $T^{\mu\nu}$, where as
in heat flow it comes from $N^{\mu}$.

\subsubsection{Shear and bulk viscous coefficients}

In both the terms of Eq.(\ref{Eq40}), $\langle\tilde{p}^{\mu}\tilde{p}^{\nu}\rangle$ can be decomposed in a traceless and a remaining part,
giving rise to $\Pi^{\mu\nu}$ a shear and a bulk part respectively.

Following this argument the shear viscous tensor comes out to be,

\begin{eqnarray}
\bar{\Pi}^{\mu\nu}=&&\Pi^{\mu\nu}-\Pi\Delta^{\mu\nu}~,\nonumber\\
=&&\sum_{k=1}^{N}\nu_k\int\frac{d^{3}|\vec{\tilde{p_k}}|}{(2\pi)^3{\omega_{p_{k}}}}\langle\langle{\tilde{p_k}^{\mu}}{\tilde{p_k}^{\nu}}\rangle\rangle f^0_k(1\pm f^0_k)\phi_k\nonumber\\
+&&\delta\omega\sum_{k=1}^{N} \  \nu_k\int\frac{d^{3}|\vec{\tilde{p_k}}|}{(2\pi)^3{\omega_{p_k}}}\frac{\langle\langle{\tilde{p_k}^{\mu}}{\tilde{p_k}^{\nu}}\rangle\rangle}{|\vec{\tilde{p_k}}|} f^0_k(1\pm f^0_k)\phi_k~.\nonumber\\
 \label{Eq47}
\end{eqnarray}
Here $\langle\langle ~ \rangle\rangle$ has been used to denote the traceless irreducible tensor,
$\langle\langle{A^{\mu}}{B^{\nu}}\rangle\rangle=\big\{\frac{1}{2}\Delta^{\mu}_{\alpha}\Delta^{\nu}_{\beta}+\frac{1}{2}\Delta^{\mu}_{\beta}\Delta^{\nu}_{\alpha}-\frac{1}{3}\Delta_{\alpha\beta}\Delta^{\mu\nu}\big\}A^{\alpha}B^{\beta}$.
Consequently, the bulk viscous part takes the following form,

\begin{eqnarray}
 \Pi=&& \sum_{k=1}^{N}\frac{\nu_k}{3}\int\frac{d^{3}|\vec{\tilde{p_k}}|}{(2\pi)^3{\omega_{p_{k}}}}\Delta_{\mu\nu}\tilde{p_k}^{\mu}\tilde{p_k}^{\nu} f^0_k(1\pm f^0_k)\phi_k\nonumber\\
    +&&\delta\omega \  \sum_{k=1}^{N}\frac{\nu_k}{3}\int\frac{d^{3}|\vec{\tilde{p_k}}|}{(2\pi)^3{\omega_{p_{k}}}}\frac{1}{|\vec{\tilde{p_k}}|}\Delta_{\mu\nu}\tilde{p_k}^{\mu}\tilde{p_k}^{\nu} f^0_k(1\pm f^0_k)\phi_k~.\nonumber\\
\label{Eq48}
\end{eqnarray}

Putting the expression of $\phi_k$ from Eq.(\ref{Eq30}) and comparing with the macroscopic definition of $\Pi^{\mu\nu}$ as follows,

\begin{equation}
 \Pi^{\mu\nu}=2\eta \langle\partial^{\mu}u^{\nu}\rangle+\zeta \Delta^{\mu\nu}\partial\cdot u~,
 \label{Eq49}
\end{equation}

we obtain following expressions of shear and bulk viscosity respectively,

\begin{eqnarray}
 \eta=&&\sum_{k=1}^{N}\frac{\nu_k\tau_k}{15T}\int\frac{d^{3}|\vec{\tilde{p_k}}|}{(2\pi)^3}\frac{|\vec{\tilde{p_k}}|^4}{\omega^2_{p_{k}}}f^0_k(1\pm f^0_k)\nonumber\\
 +&&\delta\omega\ \sum_{k=1}^{N}\frac{\nu_k\tau_k}{15T}\int\frac{d^{3}|\vec{\tilde{p_k}}|}{(2\pi)^3}\frac{|\vec{\tilde{p_k}}|^3}{\omega^2_{p_{k}}}f^0_k(1\pm f^0_k)~,
 \label{Eq50}
 \\
 \nonumber\\
 \zeta=&&\sum_{k=1}^{N}\frac{\nu_k\tau_k}{9T}\int\frac{d^{3}|\vec{\tilde{p_k}}|}{(2\pi)^3}\frac{1}{\omega^2_{p_{k}}} \big\{|\vec{\tilde{p_k}}|^2-3\omega^2_{p_k}c_s^2\big\}^2 f^0_k(1\pm f^0_k)\nonumber\\
 &&+\delta\omega\  \sum_{k=1}^{N}\frac{\nu_k\tau_k}{9T}\int\frac{d^{3}|\vec{\tilde{p_k}}|}{(2\pi)^3}\frac{1}{\omega^2_{p_{k}}}\frac{1}{|\vec{\tilde{p_k}}|}\big\{|\vec{\tilde{p_k}}|^2-3\omega^2_{p_k}c_s^2\big\}^2\nonumber\\
 &&\times f^0_k(1\pm f^0_k)~.
\label{Eq51}
 \end{eqnarray}

Clearly the second term on the right hand side of Eq.(\ref{Eq50}) and (\ref{Eq51}) is the additional term due to quasi particle excitations, over the first
term which comes from the usual kinetic theory of bare particles.

\subsubsection{Thermal conductivity}
Since, we observe only the non-equilibrium part of the heat-flow is relevant, we can obtain its analytical expression for a multi-component system,
from Eq.(\ref{Eq46}) after contracting with projection operator and hydrodynamic velocity,

\begin{eqnarray}
 \delta I^{\mu}=&&\sum_{k=1}^{N}\nu_k\int\frac{d^{3}|\vec{\tilde{p_k}}|}{(2\pi)^3{\omega_{p_{k}}}}(\omega_{p_{k}}-h_k)\langle\tilde{p_k}^{\mu}\rangle f^0_k(1\pm f^0_k)\phi_k\nonumber\\
        &&-\delta\omega\  \sum_{k=1}^{N}\nu_k\int\frac{d^{3}|\vec{\tilde{p_k}}|}{(2\pi)^3{\omega_{p_{k}}}}\frac{h_k}{|\vec{\tilde{p_k}}|} \langle\tilde{p_k}^{\mu}\rangle f^0_k(1\pm f^0_k)\phi_k~.\nonumber\\
\label{Eq52}
\end{eqnarray}

Putting the expression of $\phi_k$ from Eq.(\ref{Eq30}) and comparing with the macroscopic definition of heat flow,
\begin{equation}
 \delta I^{\mu}=\lambda T X_q^{\mu},
 \label{Eq53}
\end{equation}
we obtain the expression of thermal conductivity as follows,
\begin{eqnarray}
 \lambda=&&\sum_{k=1}^{N}\frac{\nu_k\tau_k}{3T^2}\int\frac{d^{3}|\vec{\tilde{p_k}}|}{(2\pi)^3}\frac{(\omega_{p_{k}}-h_k)^2}{\omega^2_{p_{k}}}|\vec{\tilde{p_k}}|^2 f^0_k(1\pm f^0_k)\nonumber\\
 &&-\delta\omega\  \sum_{k=1}^{N}\frac{\nu_k\tau_k}{3T^2}\int\frac{d^{3}|\vec{\tilde{p_k}}|}{(2\pi)^3}\frac{h_k(\omega_{p_{k}}-h_k)}{\omega^2_{p_{k}}}|\vec{\tilde{p_k}}| f^0_k(1\pm f^0_k)~.\nonumber\\
\label{Eq54}
 \end{eqnarray} 

The term proportional to $\delta\omega$ in Eq.(\ref{Eq54}) is the extra term here over the usual definition of $\lambda$ from kinetic theory, which
renders the quasi-particle excitation in the analytical expression of thermal conductivity.

\subsubsection{Dynamical inputs}

The relaxation times that are inverse of the collision frequencies, provide the dynamical interaction measures in the expressions of the above mentioned
transport coefficients. 
The relaxation times, corresponding to the quasi-gluons and quasi-quarks/anti-quarks, namely $\tau_g$ and 
$\tau_{q/\bar{q}}$ respectively, have been taken from \cite{Mitra1}, with updated lattice EOSs \cite{Lattice2}. The interaction cross-sections therein 
include the binary, elastic scatterings between quarks and gluons, considering the fusion processes not to be able to contribute forward peak in the 
differential cross section, leading to insignificant contribution with respect to elastic ones. The effective coupling for an interacting QCD medium, 
has been introduced following the EQPM prescription of charge renormalization \cite{Chandra3}, at finite temperature and baryon chemical potential. 
Throughout the analysis, the quark chemical potential has been taken to be $\mu_q=100$ MeV.

\section{Results and discussions}

\begin{figure}
\includegraphics[scale=0.3]{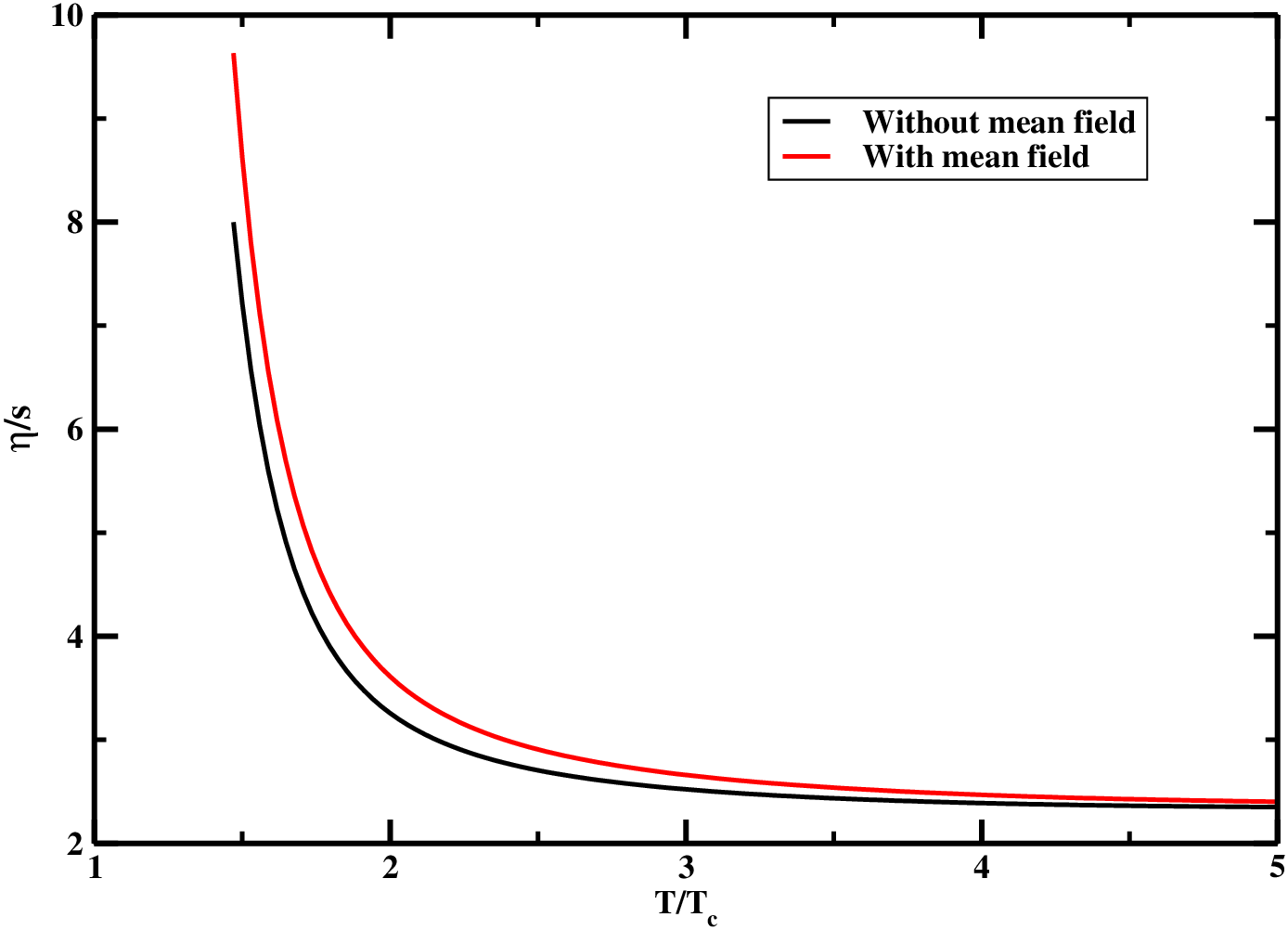}
\caption{Shear viscosity over entropy density ratio as a function of $T/T_c$ with and without mean field correction.}
\label{shear}
\end{figure}

\begin{figure}
\includegraphics[scale=0.3]{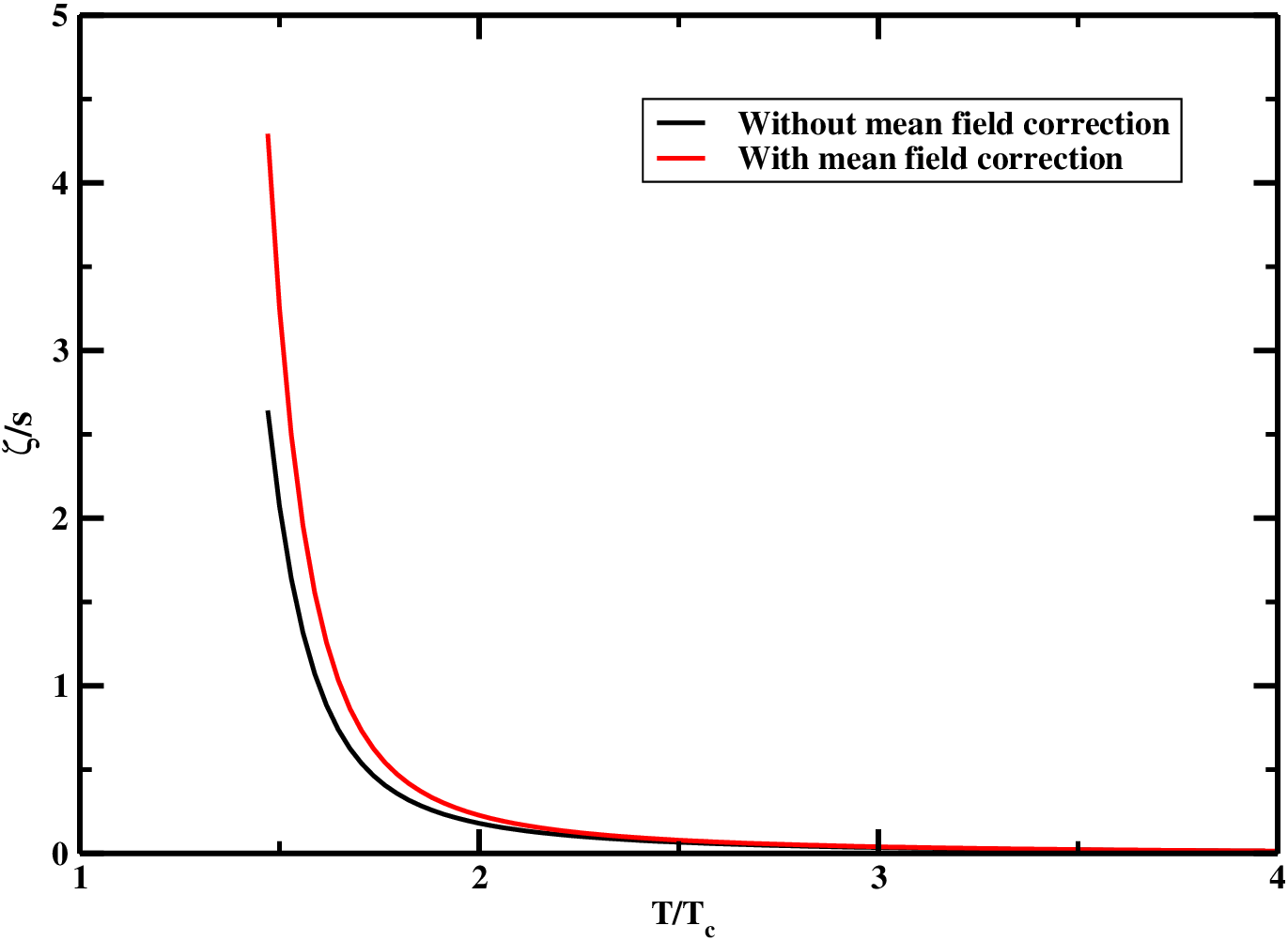}
\caption{Bulk viscosity over entropy density ratio as a function of $T/T_c$ with and without mean field correction.}
\label{bulk}
\end{figure} 

\begin{figure}
\includegraphics[scale=0.3]{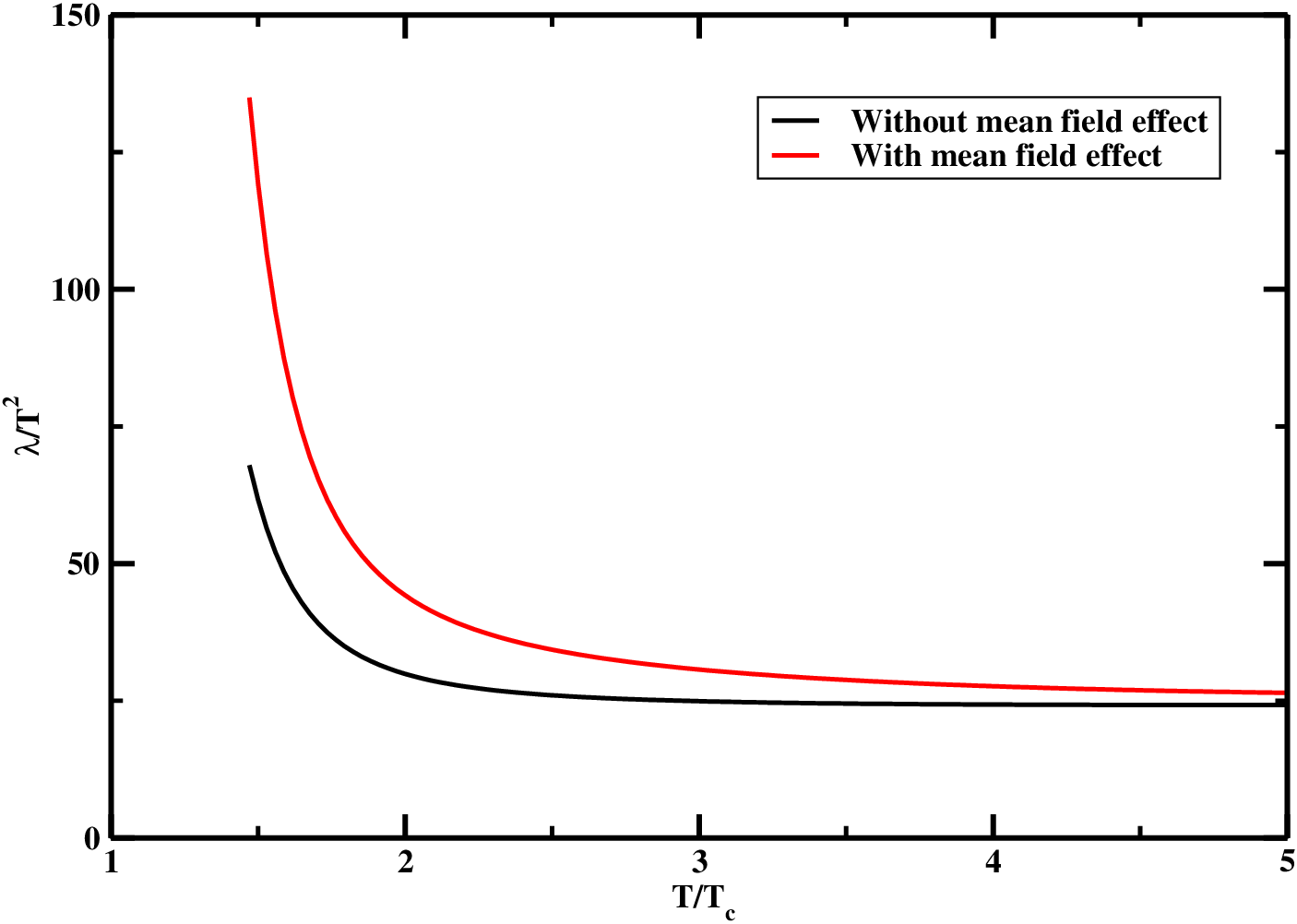}
\caption{Thermal conductivity scaled by $T^2$ as a function of $T/T_c$ with and without mean field correction.}
\label{Tcond}
\end{figure}

In this section, we have depicted the temperature dependence of shear and bulk viscous coefficients over entropy density
and thermal conductivity scaled by $T^2$, with and without the mean field corrections in Fig.(\ref{shear}), (\ref{bulk}) and
(\ref{Tcond}) respectively. As we have already discussed in our
previous work \cite{Mitra1}, the mean field corrections are of second order in gradients, which indeed appear as the 
mean field force term, $\partial_{\mu}\big\{\delta\omega u^{\mu}u^{i}\big\}$. This term being a derivative over $\delta\omega$,
which itself is  a temperature gradient of the fugacity parameter $z_{g/q}$, turns out to be second order in gradient. Since, at higher temperature regions, mostly over
$T/T_c \sim 2.5$, $z_{g/q}$ is a slowly varying function of $T$ \cite{Chandra2,Chandra3}, we can see that the mean field effects are almost negligible
in the temperature dependence of transport coefficients over $ T/T_c \sim 3$. However, below $T/T_c \sim 3$, the sharp temperature
gradient of $z_{g/q}$, makes the mean field term significant, which is consequently reflected in the temperature behavior 
of transport coefficients. In Fig.(\ref{shear}), (\ref{bulk}) and (\ref{Tcond}), the estimated values of shear viscosity, bulk viscosity and thermal conductivity
including mean field corrections, follow the same temperature trend as obtained in the previous estimation \cite{Mitra1} without considering it. However, the 
quantitative difference in the low temperature behavior of shear and bulk viscous coefficients 
as well as thermal conductivity, with and without mean field corrections, reveals the significance of the mean field term
induced by system's collective behavior, in estimating thermodynamic quantities in those temperature regions.  Therefore, although 
the mean field correction is not affecting the transport coefficients beyond $T/T_c \sim 3$, and hence neglecting those
terms are justified in explaining the high temperature behavior of the system properties, it's inclusion is essential in order to explain 
the behavior of thermodynamic parameters closer to $T_c$.

\section{Conclusions and outlook}
In conclusion, a covariant kinetic theory is developed for the hot QCD matter/QGP, employing the effective fugacity quasi-particle model (EQPM) for interacting 
hot QCD equations of state. Since the hot QCD medium effects are encoded in the gluon and quark effective fugacities as well as in the modified part of the 
dispersion relations, the mean field terms of the current effective kinetic theory also include only these fugacity parameters and their derivatives. The 
modified energy momentum tensor reproduces the hot QCD thermodynamics exactly, while respecting the thermodynamic consistency condition. The conservation laws 
are realized  in an exact way from the covariant kinetic theory while taking its appropriate moments of the relativistic transport equation. 

Interestingly, the mean field contributions induce sizable modifications to the transport coefficients of the hot QCD matter.  Be it shear viscosity, bulk 
viscosity or the thermal conductivity, the respective three momentum integrals involving the out of equilibrium part of the momentum distribution, get modified 
with an additive term proportional to $\delta \omega_{g,q}/\vert \vec{p}\vert$. Recalling that $\delta \omega_{g,q}$ is the medium modified part of the dispersion 
relation containing the collective effects of an interacting medium, we can conclude thus this additional term introduces the quasi-particle excitations in the
expressions of transport coefficient. The modifications have negligible contribution at higher temperatures ( $\geq 2.5\  T_c$) as far as the first order transport 
coefficients are concerned, however, in the vicinity of the transition temperature the mean field effects appear to be significant. This observation is in line 
with our earlier estimates for the transport coefficients such as shear and bulk viscosities, and thermal conductivity {\it etc.} \cite{Mitra1}.

The work presented in the manuscript is the first step towards developing second and third order dissipative relativistic hydrodynamics from transport theory 
with the effective fugacity quasi-particle model along with estimating the respective second and third order transport coefficients from the relativistic 
effective kinetic theory. These aspects will be taken up in immediate near future. Moreover, the electromagnetic responses of the strongly interacting medium
in presence of an electric or magnetic field, are scheduled to be explored following the line of work in \cite{Mitra2,Kurian}, within the scopes of the effective 
kinetic theory developed in the present work.

\acknowledgments
SM acknowledges Indian Institute of Technology Gandhinagar for the Institute Postdoctoral Fellowship. 
VC would like to acknowledge Department of Science and Technology  (DST), Govt. of India for INSPIRE-
Faculty Fellowship (IFA-13/PH-55) and Science and Engineering Research Board (SERB), DST for granting  
funds under Early Career Research Award (ECRA). We record our sincere gratitude to the people of India 
for their generous  help for the research in basic sciences. 

\appendix


\begin{thebibliography}{99}
  \bibitem{STAR}
  J. Adams {\it et al.} (STAR Collaboration), Nucl. Phys. A {\bf 757}, 102 (2005),
  B.~I.~Abelev {\it et al.} [STAR Collaboration],
  %``Centrality dependence of charged hadron and strange hadron elliptic flow from s(NN)**(1/2) = 200-GeV Au + Au collisions,''
  Phys.\ Rev.\ C {\bf 77} (2008) 054901.
  
  \bibitem{PHENIX}
  K. Adcox {\it et al.} (PHENIX Collaboration), Nucl. Phys.  A {\bf 757}, 184 (2005).
  
  \bibitem{PHOBOS}
  B. B. Back {\it et al.} (PHOBOS Collaboration), Nucl. Phys. A {\bf 757}, 28 (2005).
  
  \bibitem{ALICE}
  K. Aamodt {\it et al.} (The Alice Collaboration), Phys. Rev. Lett. {\bf 105}, 252302 (2010);
  Phys. Rev.  Lett. {\bf 105}, 252301 (2010); Phys. Rev. Lett. {\bf 106}, 032301 (2011),
  J.~Adam {\it et al.} [ALICE Collaboration],
  %``Higher harmonic flow coefficients of identified hadrons in Pb-Pb collisions at $\sqrt{s_{\rm NN}}$ = 2.76 TeV,''
  JHEP {\bf 1609} (2016) 164.
  
  \bibitem{Lattice1}
  A.~Bazavov {\it et al.},
  %``Equation of state and QCD transition at finite temperature,''
  Phys.\ Rev.\ D {\bf 80} (2009) 014504.
  
  \bibitem{Lattice2}
  A.~Bazavov {\it et al.} [HotQCD Collaboration],
  %``Equation of state in ( 2+1 )-flavor QCD,''
  Phys.\ Rev.\ D {\bf 90} (2014) 094503.
  
  \bibitem{Lattice3}
  S.~Borsanyi, G.~Endrodi, Z.~Fodor, S.~D.~Katz, S.~Krieg, C.~Ratti and K.~K.~Szabo,
  %``QCD equation of state at nonzero chemical potential: continuum results with physical quark masses at order $mu^2$,''
  JHEP {\bf 1208} (2012) 053.
  
  \bibitem{Lattice4}
   S.~Borsanyi, Z.~Fodor, C.~Hoelbling, S.~D.~Katz, S.~Krieg and K.~K.~Szabo,
  %``Full result for the QCD equation of state with 2+1 flavors,''
  Phys.\ Lett.\ B {\bf 730} (2014) 99.
  
  \bibitem{HTLpt}
  N.~Haque, A.~Bandyopadhyay, J.~O.~Andersen, M.~G.~Mustafa, M.~Strickland and N.~Su,
  %``Three-loop HTLpt thermodynamics at finite temperature and chemical potential,''
  JHEP {\bf 1405} (2014) 027.
  
  \bibitem{AMY1}
  P.~B.~Arnold, G.~D.~Moore and L.~G.~Yaffe,
  %``Transport coefficients in high temperature gauge theories. 1. Leading log results,''
  JHEP {\bf 0011} (2000) 001.
  
  \bibitem{AMY2}
  P.~B.~Arnold, G.~D.~Moore and L.~G.~Yaffe,
  %``Transport coefficients in high temperature gauge theories. 2. Beyond leading log,''
  JHEP {\bf 0305} (2003) 051.
  
  \bibitem{Chen}
  J.~W.~Chen, J.~Deng, H.~Dong and Q.~Wang,
  %``Shear and bulk viscosities of a gluon plasma in perturbative QCD: Comparison of different treatments for the gg↔ggg process,''
  Phys.\ Rev.\ C {\bf 87} (2013) 024910,
  J.~W.~Chen, H.~Dong, K.~Ohnishi and Q.~Wang,
  %``Shear Viscosity of a Gluon Plasma in Perturbative QCD,''
  Phys.\ Lett.\ B {\bf 685} (2010) 277.
  
  \bibitem{Arnold}
  P.~B.~Arnold, C.~Dogan and G.~D.~Moore,
  %``The Bulk Viscosity of High-Temperature QCD,''
  Phys.\ Rev.\ D {\bf 74} (2006) 085021.
  
  \bibitem{Xu}
  Z.~Xu and C.~Greiner,
  %``Shear viscosity in a gluon gas,''
  Phys.\ Rev.\ Lett.\  {\bf 100} (2008) 172301.
  
  \bibitem{Peshier}
   A.~Peshier, B.~Kampfer, O.~P.~Pavlenko and G.~Soff,
  %``An Effective model of the quark - gluon plasma with thermal parton masses,''
  Phys.\ Lett.\ B {\bf 337} (1994) 235,
   A.~Peshier, B.~Kampfer, O.~P.~Pavlenko and G.~Soff,
  %``A Massive quasiparticle model of the SU(3) gluon plasma,''
  Phys.\ Rev.\ D {\bf 54} (1996) 2399,
  A.~Peshier, B.~Kampfer and G.~Soff,
  %``The Equation of state of deconfined matter at finite chemical potential in a quasiparticle description,''
  Phys.\ Rev.\ C {\bf 61} (2000) 045203, 
  A.~Peshier, B.~Kampfer and G.~Soff,
  %``From QCD lattice calculations to the equation of state of quark matter,''
  Phys.\ Rev.\ D {\bf 66} (2002) 094003.
  
  \bibitem{Rebhan}
  A.~Rebhan and P.~Romatschke,
  %``HTL quasiparticle models of deconfined QCD at finite chemical potential,''
  Phys.\ Rev.\ D {\bf 68} (2003) 025022.
  
  \bibitem{Bluhm1}
  M.~Bluhm, B.~Kampfer and G.~Soff,
  %``Quasi-particle model of strongly interacting matter,''
  J.\ Phys.\ G {\bf 31} (2005) S1151.
  
  \bibitem{Thaler}
  M.~A.~Thaler, R.~A.~Schneider and W.~Weise,
  %``Quasiparticle description of hot QCD at finite quark chemical potential,''
  Phys.\ Rev.\ C {\bf 69} (2004) 035210.
  
  \bibitem{Szabo}
  K.~K.~Szabo and A.~I.~Toth,
  %``Quasiparticle description of the QCD plasma, comparison with lattice results at finite T and mu,''
  JHEP {\bf 0306} (2003) 008.
  
  \bibitem{Bannur}
  V.~M.~Bannur,
  %``Quasi-particle model for QGP with nonzero densities,''
  JHEP {\bf 0709} (2007) 046,
  V.~M.~Bannur,
  %``Comments on quasiparticle models of quark-gluon plasma,''
  Phys.\ Lett.\ B {\bf 647} (2007) 271.
  
  \bibitem{Chandra1}
  V.~Chandra and V.~Ravishankar,
  %``Viscosity and thermodynamic properties of QGP in relativistic heavy ion collisions,''
  Eur.\ Phys.\ J.\ C {\bf 59} (2009) 705.
  
  \bibitem{Chandra2}
  V.~Chandra and V.~Ravishankar,
  %``Quasi-particle model for lattice QCD: Quark-gluon plasma in heavy ion collisions,''
  Eur.\ Phys.\ J.\ C {\bf 64} (2009) 63.
  
  \bibitem{Chandra3}
  V.~Chandra and V.~Ravishankar,
  %``A quasi-particle description of $(2+1)$- flavor lattice QCD equation of state,''
  Phys.\ Rev.\ D {\bf 84} (2011) 074013.
  
  \bibitem{Jeon} 
  S.~Jeon,
  %``Hydrodynamic transport coefficients in relativistic scalar field theory,''
  Phys.\ Rev.\ D {\bf 52}, 3591 (1995).
  
  \bibitem{Dusling} 
  K.~Dusling and T.~Schäfer,
  %``Bulk viscosity, particle spectra and flow in heavy-ion collisions,''
  Phys.\ Rev.\ C {\bf 85}, 044909 (2012).
  
  \bibitem{Gorenstein}
  M.~I.~Gorenstein and S.~N.~Yang,
  %``Gluon plasma with a medium dependent dispersion relation,''
  Phys.\ Rev.\ D {\bf 52} (1995) 5206.
  
  \bibitem{Tinti}
  L.~Tinti, A.~Jaiswal and R.~Ryblewski,
  %``Quasiparticle second-order viscous hydrodynamics from kinetic theory,''
  Phys.\ Rev.\ D {\bf 95} (2017) no.5,  054007.
  
  \bibitem{Bluhm2}
  M.~Bluhm, B.~Kampfer and K.~Redlich,
  %``Bulk and shear viscosities of the gluon plasma in a quasiparticle description,''
  Phys.\ Rev.\ C {\bf 84} (2011) 025201.
  
  \bibitem{Kapusta1}
  M.~Albright and J.~I.~Kapusta,
  %``Quasiparticle Theory of Transport Coefficients for Hadronic Matter at Finite Temperature and Baryon Density,''
  Phys.\ Rev.\ C {\bf 93} (2016) no.1,  014903.
  
  \bibitem{Kapusta2}
  P.~Chakraborty and J.~I.~Kapusta,
  %``Quasi-Particle Theory of Shear and Bulk Viscosities of Hadronic Matter,''
  Phys.\ Rev.\ C {\bf 83} (2011) 014906.
  
  \bibitem{Greco1}
  S.~Plumari, A.~Puglisi, F.~Scardina and V.~Greco,
  %``Shear Viscosity of a strongly interacting system: Green-Kubo vs. Chapman-Enskog and Relaxation Time Approximation,''
  Phys.\ Rev.\ C {\bf 86} (2012) 054902.
  
  \bibitem{Greco2}
  A.~Puglisi, S.~Plumari and V.~Greco,
  %``Shear viscosity η to electric conductivity σ$_{el}$ ratio for the quark–gluon plasma,''
  Phys.\ Lett.\ B {\bf 751} (2015) 326.
  
  \bibitem{Alqahtani1}
  M.~Alqahtani, M.~Nopoush and M.~Strickland,
  %``Quasiparticle equation of state for anisotropic hydrodynamics,''
  Phys.\ Rev.\ C {\bf 92} (2015) no.5,  054910.
  
  \bibitem{Alqahtani2}
  M.~Alqahtani, M.~Nopoush and M.~Strickland,
  %``Quasiparticle anisotropic hydrodynamics for central collisions,''
  Phys.\ Rev.\ C {\bf 95} (2017) no.3,  034906.
  
  \bibitem{Combridge}
  B.~L.~Combridge, J.~Kripfganz and J.~Ranft,
  %``Hadron Production at Large Transverse Momentum and QCD,''
  Phys.\ Lett.\  {\bf 70B} (1977) 234.
  
  \bibitem{Mitra1}
  S.~Mitra and V.~Chandra,
  %``Transport coefficients of a hot QCD medium and their relative significance in heavy-ion collisions,''
  Phys.\ Rev.\ D {\bf 96} (2017) no.9,  094003.
  
  \bibitem{Mitra2}
  S.~Mitra and V.~Chandra,
  %``Thermal relaxation, electrical conductivity, and charge diffusion in a hot QCD medium,''
  Phys.\ Rev.\ D {\bf 94} (2016) no.3,  034025.
  
  \bibitem{Kurian}
  M.~Kurian and V.~Chandra,
  %``Effective description of hot QCD medium in strong magnetic field and longitudinal conductivity,''
  Phys.\ Rev.\ D {\bf 96} (2017) 114026.
   
\end{thebibliography}
\end{document}